\documentclass[pra,onecolumn]{revtex4}%
\usepackage{epsfig}
\usepackage{amsmath}
\usepackage{graphicx}
\usepackage{bm}
\usepackage{amsfonts}
\usepackage{amssymb}

\begin{document}
\title{Extremely local electric field enhancement and light confinement in dielectric waveguide}
\author{Qijing Lu, $^{1}$ Fang-Jie Shu,$^{2}$ Chang-Ling Zou.$^{1,*}$}
\address{
$^1$ Key Lab of Quantum Information, University of Science and Technology of
China, Hefei 230026, P. R. China \\
$^2$ Department of Physics, Shangqiu Normal University, Shangqiu 476000, P. R. China}
 \email{clzou321@ustc.edu.cn}
\date{\today}

\begin{abstract}
The extremely local electric field enhancement and light confinement
is demonstrated in dielectric waveguide with corner and gap geometry.
The numerical results reveal the local electric field enhancement
in the vicinity of the apex of fan-shaped waveguide. Classical electromagnetic
theory predicts that the field enhancement and confinement abilities
increase with decreasing radius of rounded corner ($r$) and gap ($g$),
and show singularity for infinitesimal $r$ and $g$. For practical
parameters with $r=g=10\,\mathrm{nm}$, the mode area of opposing
apex-to-apex fan-shaped waveguides can be as small as $4\times10^{-3}A_{0}$
($A_{0}=\lambda^{2}/4$), far beyond the diffraction limit. This way
of breaking diffraction limit with no loss outperforms plasmonic waveguides,
where light confinement is realized at the cost of huge intrinsic
loss in the metal. Furthermore, we propose a structure with dielectric
bow-tie antenna on a silicon-on-insulator waveguide, whose field enhancement
increases by one order. The lossless dielectric corner and gap structures
offer an alternative method to enhance the light-matter interaction
without metal nano-structure, and will find applications in quantum
electrodynamics, sensors and nano-particle trapping.
\end{abstract}

\maketitle

\section{Introduction}

Strong light confinement in photonic devices can enhance the light
field intensity, and then lead to very strong light-matter interaction.
The enhanced light field is essential for a wide range of applications,
such as quantum electrodynamics \cite{1,2}, nonlinear optical effect
\cite{3}, quantum optomechanics \cite{4}, optical sensors \cite{sensor}
and nano-optical tweezers \cite{tweezer}. Therefore, the surface
plasmon in metal nanostructures is attracting more and more attentions
for its unique ability to confine light in the deep subwavelength
scale. Extreme strong field enhancements in surface plasmon are mainly
attributed to the very small geometry size \cite{sppnano,nanolaser},
sharp corners \cite{wedge-exp,wedge-prl,wedge-qiu} and nanoscale
gaps \cite{gap1,bowtie-1,bowtie-2}, which has been stated in Ref.\cite{enhance}.
However, intrinsic loss due to internal damping of radiation in metal
limits the explosion of practical applications.

For the case of dielectric, light is confined in the wavelength-scale
photonic devices, with weak evanescent field interacting with outside.
Great efforts have been dedicated to enhance the light field intensity
by engineering the dielectric structure. In 2004, Almeida et al. have
proposed the dielectric slot waveguide structure, and demonstrated
a very effective way to enhance the light field in the void \cite{slot-1,slot-2}.
Very recently, three-dimensional photonic crystal structure has been
proposed to guide the light in the wedge-like waveguide with field
enhanced at the apex \cite{pcedge}. Actually, the electric field
enhancement in the vicinity of the dielectric corner is well known
in electrostatics \cite{corner2}. While in the studies of electromagnetic
waves, the corner effect is noticed because it may lead to diffraction
at corners and computation difficulties in rectangle-cross-section
waveguides \cite{corner1,corner3,corner4}.

In this paper, we numerically demonstrated the extremely local electric
field enhancement and light confinement in dielectric waveguide by
the corner and gap structures. The waveguide with fan-shaped cross-section
is proposed to utilize the corner effect \cite{corner2,corner1,corner3,corner4}
to enhance the local electric field, while the opposing apex-to-apex
fan-shaped waveguides are studied to apply the gap effect \cite{slot-1,slot-2}.
It is shown that the mode area can be as small as $4\times10^{-3}$
of diffraction limited area for practical geometry, superior to the
confinement in hybrid surface plasmon waveguides \cite{Oulton}. Furthermore,
we proposed a structure with dielectric bow-tie antenna on a silicon-on-insulator
(SOI) waveguide to enhance the amplitude of electric field by one
order. It is worth noting that the very sharp corner and closed gap
give rise to singularity behavior of local field enhancement, which
calls for theoretical efforts to study the quantum effects of dielectric
at atomic scale \cite{quantum-1,quantum-2,JPCC}.

\section{Field enhancement by corner}

\begin{figure}
\label{fig1}\centerline{ \includegraphics[width=0.7\columnwidth]{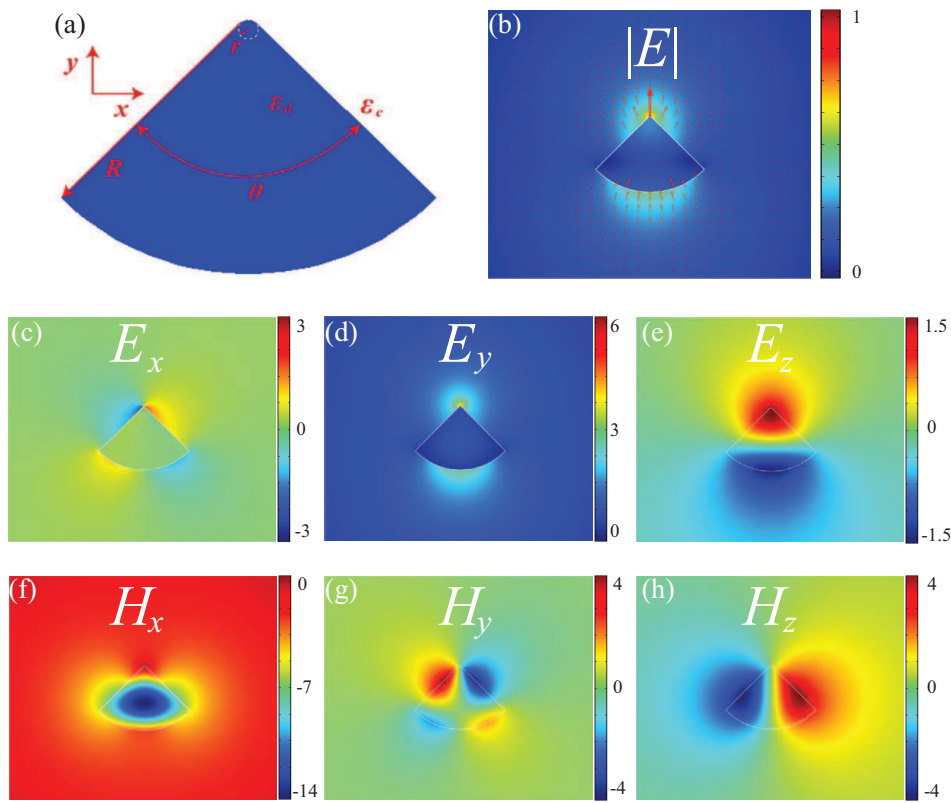}}
\caption{(a) Schematic illustration of the fan-shaped semiconductor waveguide
with a sharp corner. (b) $|E|$ distribution for the fundamental wedge
mode. Red arrows indicate the direction of the electric field, which
are mainly along the symmetric axis of the fan. (c)-(h) Distributions
of electric and magnetic field components $E_{x}$, $E_{y}$, $E_{z}$,
$H_{x}$, $H_{y}$, $H_{z}$ for the wedge mode. Here, $R=280$ nm,
$r=10$ nm and $\theta=90$$^{\circ}$. }
\end{figure}

First of all, we proposed a fan-shaped waveguide to study the local
field enhancement at the dielectric corner. As schematically illustrated
in Fig. 1(a), the radius and angle of the fan are $R$ and $\theta$,
respectively. Based on practical situations, the apex of fan is rounded
corner with radius of curvature $r$. In our model, the waveguide
is made by silicon and embedded in air, with permittivities of waveguide
and air cladding being $\epsilon_{d}=12.25$ and $\epsilon_{c}=1$
at the working wavelength $\lambda=1550\,\mathrm{nm}$. All the following
results are obtained from the classical electromagnetic theory, which
are solved by the Finite Element Method numerically, with commercial
available software (COMSOL Multiphysics 4.3).

Figure 1(b) shows the electric field ($|E|$) distribution of the
fundamental wedge waveguide mode with $R=280$ nm, $r=10$ nm and
$\theta=90$$^{\circ}$. Clearly, electric field is greatly enhanced
in the vicinity of the apex of fan-shaped waveguide. The polarization
of the wedge mode is mainly along the axis of symmetry of fan (i.e.
$y$ axis), as indicated by the red arrows. Electric and magnetic
field components along $x,$ $y,$ $z$ directions are displayed in
Figs. 1(c)-(h), where we can find that only electric field component
$E_{y}$ is enhanced in the vicinity of the apex of corner, showing
local electric field enhancement.

\begin{figure}
\centerline{\includegraphics[width=0.8\columnwidth]{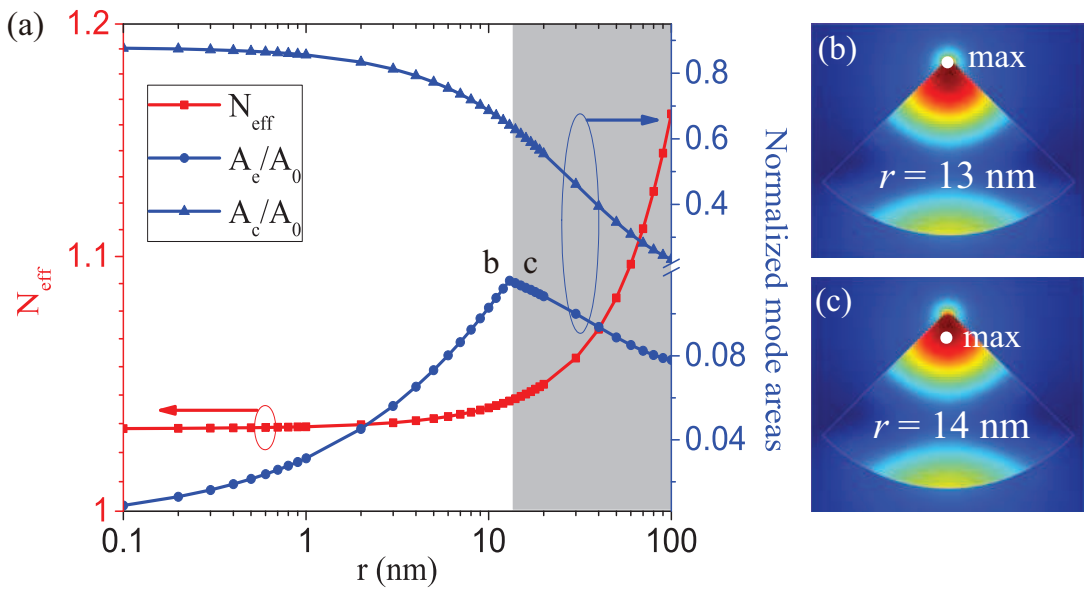}}
\caption{(a) Dependence of the effective mode index and normalized mode areas
on the radius of the curvature $r$. (b)-(c) Electric energy density
$W_{e}$ when $r=13\,\mathrm{nm}$ and $r=14\,\mathrm{nm}$, respectively.
White dots denote the $max\left\{ W_{e}(\mathbf{r})\right\} $. Here,
$R=280\:\mathrm{nm}$ and $\theta=90$$^{\circ}$.}
\label{fig2}
\end{figure}
 In the following, we'll analyze the properties of the wedge modes
in detail. Two types of effective mode area of waveguide are introduced
to measure the local field enhancement \cite{modearea}
\begin{equation}
A_{e}=\frac{\int\int_{all}W_{e}(\mathbf{r})d^{2}\mathbf{r}}{max\left\{ W_{e}(\mathbf{r})\right\} },\label{eq:1}
\end{equation}
and spatial light confinement \cite{modearea}
\begin{equation}
A_{c}=\frac{\left[\int\int_{all}W_{e}(\mathbf{r})d^{2}\mathbf{r}\right]^{2}}{\int\int_{all}[W_{e}(\mathbf{r})]^{2}d^{2}\mathbf{r}}.\label{eq:3}
\end{equation}

\noindent \begin{flushleft}
Here $W_{e}(\mathbf{r})$ is the electric field energy density,
\par\end{flushleft}

\begin{equation}
W_{e}(\mathbf{r})=\frac{1}{2}\varepsilon(\mathbf{r})|E(\mathbf{r})|^{2},\label{eq:2}
\end{equation}

\noindent with $|E(\mathbf{r})|$ and $\varepsilon(\mathbf{r})$ being
electric field and dielectric permittivity, respectively. $A_{e}$
is more sensitive to $max\left\{ W_{e}(\mathbf{r})\right\} $ , and
is usually applied to quantify local field enhancement and estimate
the enhanced spontaneous emission due to the Purcell effect \cite{modearea}.
$A_{c}$ characterizes the spatial extent of the field, which is not
sensitive to the local field enhancement.

In Fig. 2(a), the effective mode index $N_{eff}$ and normalized mode
areas $A_{e,c}/A_{0}$ are plotted against the radius of rounded corner
$r$, where $A_{0}=\lambda^{2}/4$ denotes the diffraction limited
area of vacuum. When $r$ decreases from $100\,\mathrm{nm}$ to $0.1\,\mathrm{nm}$,
$N_{eff}$ decreases and $A_{c}$ increases monotonously with opposite
trend. Note that $N_{eff}$ also represents the confinement of light
field, since strong confinement means more energy confined in dielectric
and leads to larger effect mode index. For large $r$, larger $r$
means larger dielectric cross-section area, so the confinement changes
a lot with $r$. At very small $r$, the dielectric cross-section
area does not change much, so $N_{eff}$ and $A_{c}$ show saturation
when $r<1\,\mathrm{nm}$.

On the contrary, $A_{e}$ increases with decreasing $r$ in the shadow
region with $r>13.5\,\mathrm{nm}$, similar to $A_{c}$, since the
cross-section area of dielectric increases. But it decreases with
$r$ when $r<13.5\,\mathrm{nm}$, with an inflection at $r\approx13.5\,\mathrm{nm}$.
Carefully comparing the field distributions in the two regimes (see
Figs. 2(b) and 2(c) with $r=13\,\mathrm{nm}$ and $r=14\,\mathrm{nm}$),
we found that the location of the maximum of the electric energy density
($\mathrm{max\{W_{e}(\mathbf{r})\}}$) is escaping from the core of
the waveguide to the corner as $r$ decreases. For $r<13.5\,\mathrm{nm}$,
$\mathrm{max\{W_{e}(\mathbf{r})\}}$ at the corner increases with
decreasing $r$, which shows very strong local electric field enhancement
due to the corner structure. Compared to the case of $r=10\,\mathrm{nm}$,
the mode area $A_{e}$ for the sharper corner with $r=0.1\,\mathrm{nm}$
has been reduced to about $1/10$.

\begin{figure}
\centerline{\includegraphics[width=0.6\columnwidth]{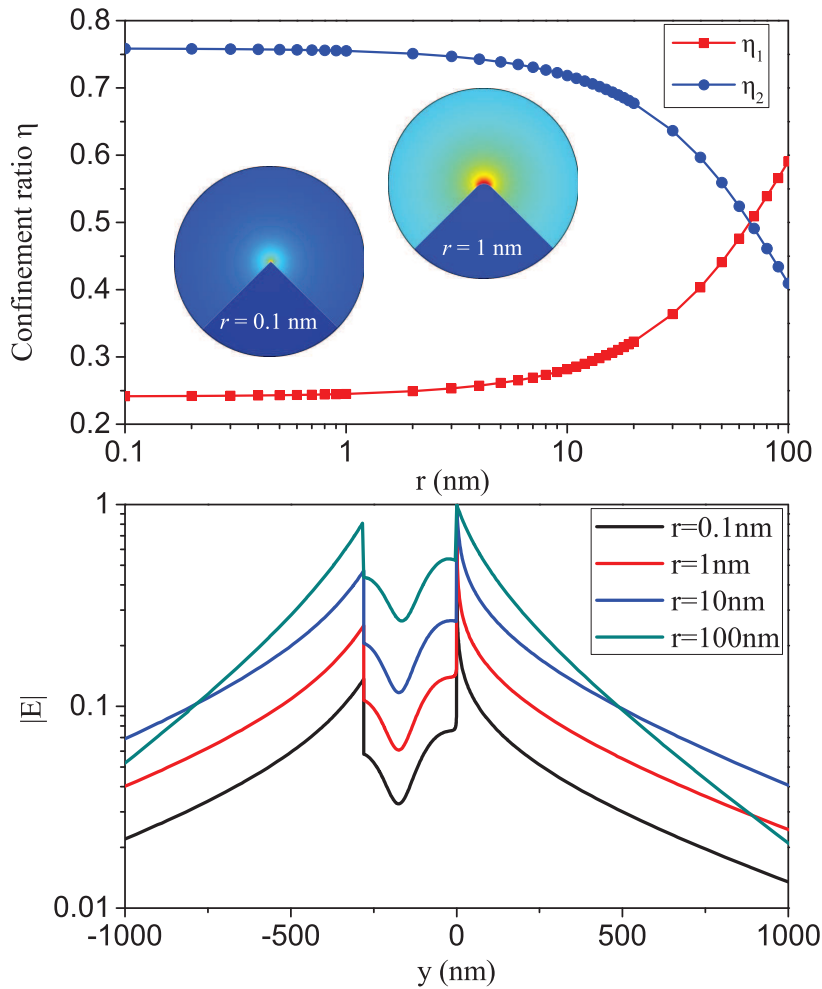}}
\caption{(a) Electric energy confinement factors in Si ($\eta_{1}$) and air
($\eta_{2}$), respectively. Inset: enlarged view of the $|E|$ distributions
of the wedge mode for $r=0.1$ nm and 1 nm, respectively. (b) Normalized
electric density along $y$ axis when $r=0.1$ nm, $r=1$ nm, $r=10$
nm, $r=100$ nm, respectively.}

\label{fig3}
\end{figure}

In Fig. 3(a), the field confinement is further studied by the energy
confinement ratios in silicon ($\eta_{1}$) and air ($\eta_{2}$),
where $\eta_{1(2)}$ is defined as the ratio of the electric energy
in the silicon (air) to the total electric energy and satisfies $\eta_{1}+\eta_{2}=1$.
The behavior of $\eta_{1}$ against $r$ agrees well with that of
$N_{eff}$ (Fig. 2(a)), since more energy in silicon leads to larger
$N_{eff}$. The curves of $A_{c}$, $N_{eff}$ and $\eta_{1}$ indicate
that the dielectric corner does not lead to stronger field confinement,
but just local electric field enhancement. Insets of Fig. \ref{fig3}(a)
are the electric field energy densities for $r=0.1\,\mathrm{nm}$
and $r=1\,\mathrm{nm}$, which demonstrate that the field enhancement
is more localized to the corner for smaller $r$. In Fig. \ref{fig3}(b),
the normalized electric fields ($|E|$) along $y$ axis for various
$r$ are shown. The field profiles are similar except the enhancement
at the corner ($y=0$), where $|E|$ around the apex decays more drastically
for smaller $r$.

\begin{figure}
\centerline{\includegraphics[width=0.6\columnwidth]{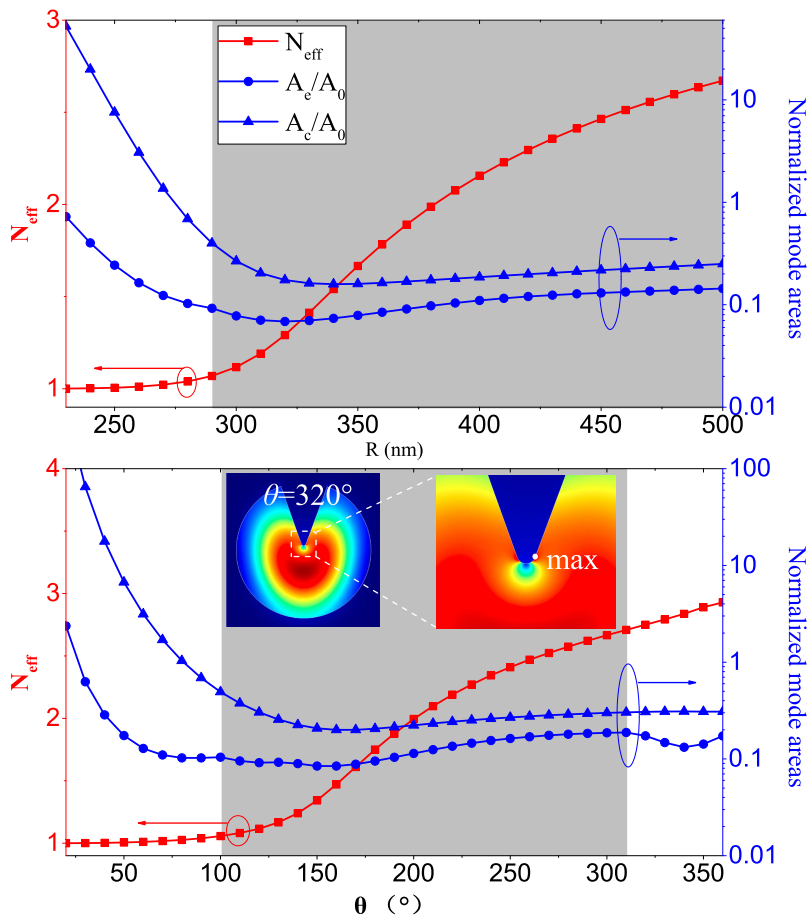}}
\caption{Dependence of the effective mode index and normalized mode areas on
$R$ (a) and $\theta$ (b), respectively. Insets in Fig. \ref{fig4}(b)
are electric energy density $W_{e}$ for $\theta=320{^\circ}$ and
enlarged view of the $max\left\{ W_{e}(\mathbf{r})\right\} $ for
$\theta=320\text{\ensuremath{{^\circ}}}$.}
\label{fig4}
\end{figure}

In Fig. \ref{fig4}, the corner effect is studied for different waveguide
size $R$ and corner angle $\theta$. For very small $R$ or $\theta$,
$A_{e}$ and $A_{c}$ are both very large due to the weak confinement
of light allowing for the small cross-section area of waveguide. When
$R$ or $\theta$ increases, $N_{eff}$ increases monotonously, while
$A_{e}$ and $A_{c}$ decrease first and then slowly increase as the
cross-section area increases. The shadow regions in Fig. 4 correspond
to the $\mathrm{max[\{W(\mathbf{r})\}}$ located inside the dielectric,
while the white regions correspond to the local electric field enhanced
around the apex of corner. For increasing $\theta$, the wedge mode
of fan-shaped waveguide is converted to the channel mode, as shown
by the insets in Fig. \ref{fig4}(b).

\section{Field enhancement by gap}

\begin{figure}
\centerline{\includegraphics[width=0.8\columnwidth]{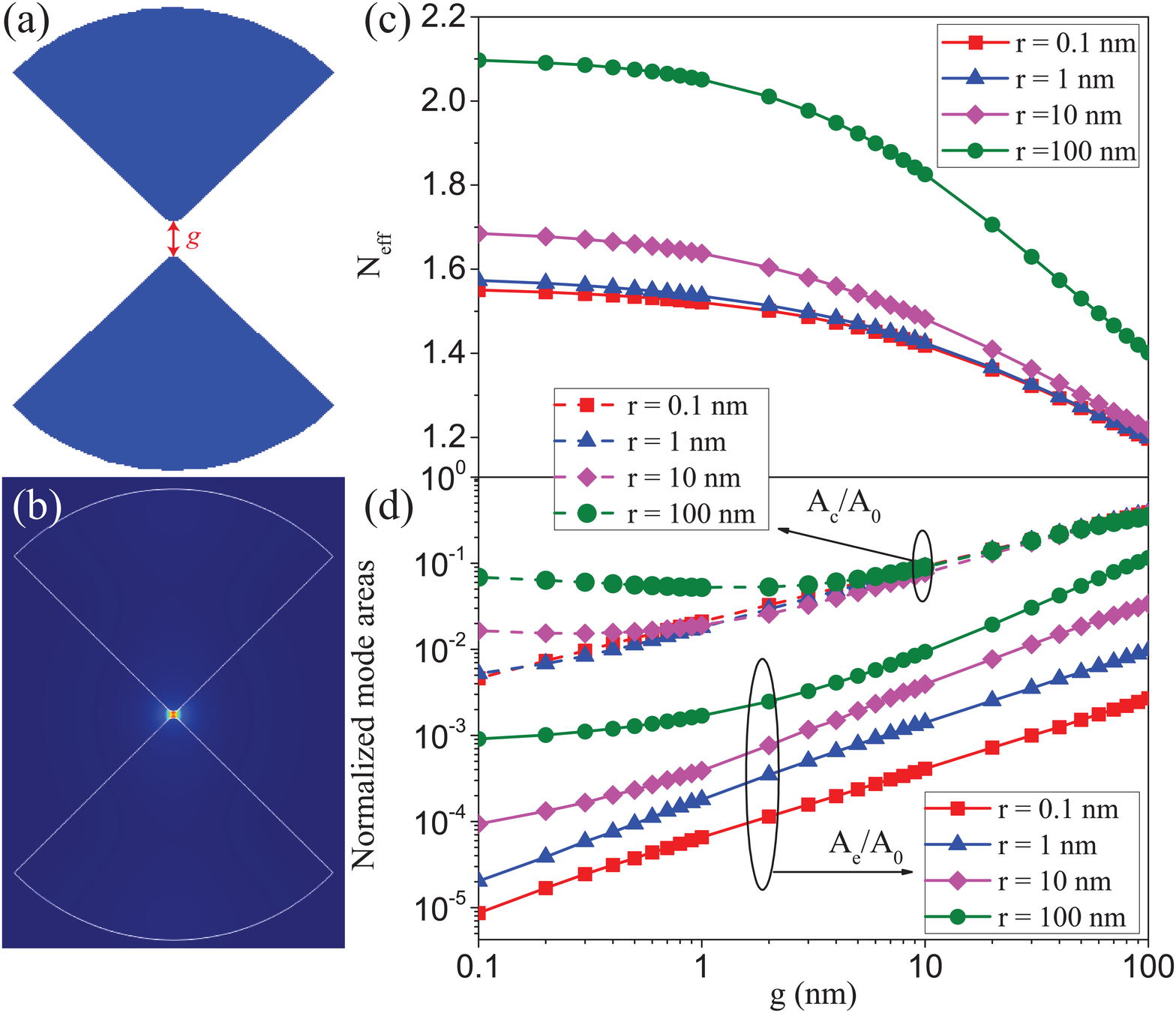}}
\caption{(a) Schematic of the bow-tie waveguide. (b) Electric energy density
$W_{e}$ of the coupled mode in the bow-tie waveguide for $R=280$
nm, $r=10$ nm, $\theta=90$$^{\circ}$ and $g=10$ nm. $N_{eff}$
(c) and normalized mode areas (d) of the coupled mode as a function
of $g$, for $r=0.1$ nm, $r=1$ nm, $r=10$ nm, $r=100$ nm, respectively.}
\label{fig5}
\end{figure}

Gap effect for light enhancement and confinement, caused by large
discontinuity of the electric field at high index-contrast interfaces,
was studied in Refs. \cite{slot-1,slot-2}, where low-index slot is
embedded between two high-index rectangle waveguides. Similar to dielectric
slot structures, we can expect further field enhancement in the double
fan waveguide posited as apex-to-apex with an air gap $g$ (Fig. \ref{fig5}(a)).
As an example shown in Fig. \ref{fig5}(b), we see the very strong
field confinement at the gap between the apexes with $g=10\,\mathrm{nm}$.

Dependences of the modal characteristics of the gap mode on $g$ are
displayed in Figs. 5(c) and 5(d). $N_{eff}$ increases monotonously
for decreasing $g$, because the low-index air surrounding the apex
is replaced by high-index dielectric. The trends of $N_{eff}$ with
different $r$ are the same, and show saturation for very small gap
$g<1\,\mathrm{nm}$. For a fixed $g$, sharper corner gives rise to
weaker light confinement in dielectric, which is consistent with the
results of single fan-shaped waveguide (Fig. \ref{fig2}(a)). For
the mode areas (\ref{fig5}(d)), both $A_{e}$ and $A_{c}$ show great
reduction when reducing the gap. These reveal that the slot structure
leads to strong spatial confinement, as well as local field enhancement.
However, $A_{c}$ is saturated when $g<1\,\mathrm{nm}$ for $r\geq10\,\mathrm{nm}$,
while $A_{e}$ is saturated only for $r=100\,\mathrm{nm}$. This indicates
that the gap effect of apex-to-apex structure also depends on the
sharpness of corners.

Due to the combined effects of corner and gap structure, we found
that $A_{e}$ can be as small as $10^{-5}A_{0}$ for $g=0.1$ nm and
$r=0.1$ nm. Compared to the best light confinement $A_{e}^{'}\approx5\times10^{-3}A_{0}$
in the hybrid dielectric-metal surface plasmon waveguide \cite{Oulton},
$A_{e}$ is $500$ times smaller. It is notable that the maximum of
electric field is always located at the corner of waveguide, which
is very suitable for light to interact with matters outside the dielectric.
The extreme light confinement and local electric field enhancement
in the dielectric corner and gap would benefit various applications,
such as the waveguide quantum electrodynamics, nanoparticle trapping,
and bio-sensors.

\section{Quantum limitations}

From the tendency in Fig. 5(d), we can predict that the smaller effective
mode area can be obtained by smaller gap or sharper corners. However,
when the size of gap or corner structures is reduced to the atomic
scale (sub-nanometer), the classical electromagnetic theory will break
down and the quantum mechanical effects appear in three aspects:

(1) The quantum size effect. The object with geometry smaller than
atomic scale is meaningless. The sub-nanometer-scale geometry should
be treated as atom cluster instead of regular corner or straight boundary.
In the atomic scale, the optical response of atomic clusters can not
be deduced by the classical dielectric constant of bulk.

(2) The non-local effect. The atomic-scale wave-function of electrons
can spread out from the boundary. Therefore, the response of the electron
to electromagnetic wave is no longer local.

(3) The quantum tunneling. When the gap between two apexes is sub-nanometer,
the electrons may tunnel across the gap. Therefore, the electromagnetic
field theory for dielectric with zero conductivity is not valid.

These quantum effects in metal nanostructures have been studied extensively
recently both in experiments and theoretical studies \cite{quantum-1,quantum-2,JPCC}.
In theory, the optical responses of those structures at sub-nanometer
scale are studied under quantum model with time-dependent density
function theory \cite{quantum-1}, or under semiclassical model with
modified electromagnetic theory by including hydrodynamic description
of conducting electrons \cite{quantum-2,JPCC}. It is demonstrated
that the singular behavior is avoided in those theoretical models
and experiments. Therefore, we could also expect the singular behavior
to be removed by corrected models. However, the property of electrons
in dielectric is significantly different from that in metal, so new
models should be developed to study the optical response of the atomic
scale dielectric structures.

Although the light confinement and local field enhancement may be
limited by quantum effects, the apex-to-apex dielectric structure
studied here is still superior to the plasmon waveguides: (a) Optical
modes in the apex-to-apex fan waveguides are lossless, which is an
incomparable advantage over plasmonic waveguides where light confinement
is realized at the cost of huge intrinsic loss in the metal. (b) Even
for the case of $g=10\,\mathrm{nm}$ and $r=10\,\mathrm{nm}$ where
the classical Maxwell equations are valid, we get $A_{e}\approx4\times10^{-3}A_{0}$
which is smaller than that of hybrid plasmonic waveguide \cite{Oulton}.

\section{Dielectric Bow-tie Antenna}

\begin{figure}
\centerline{\includegraphics[width=0.65\columnwidth]{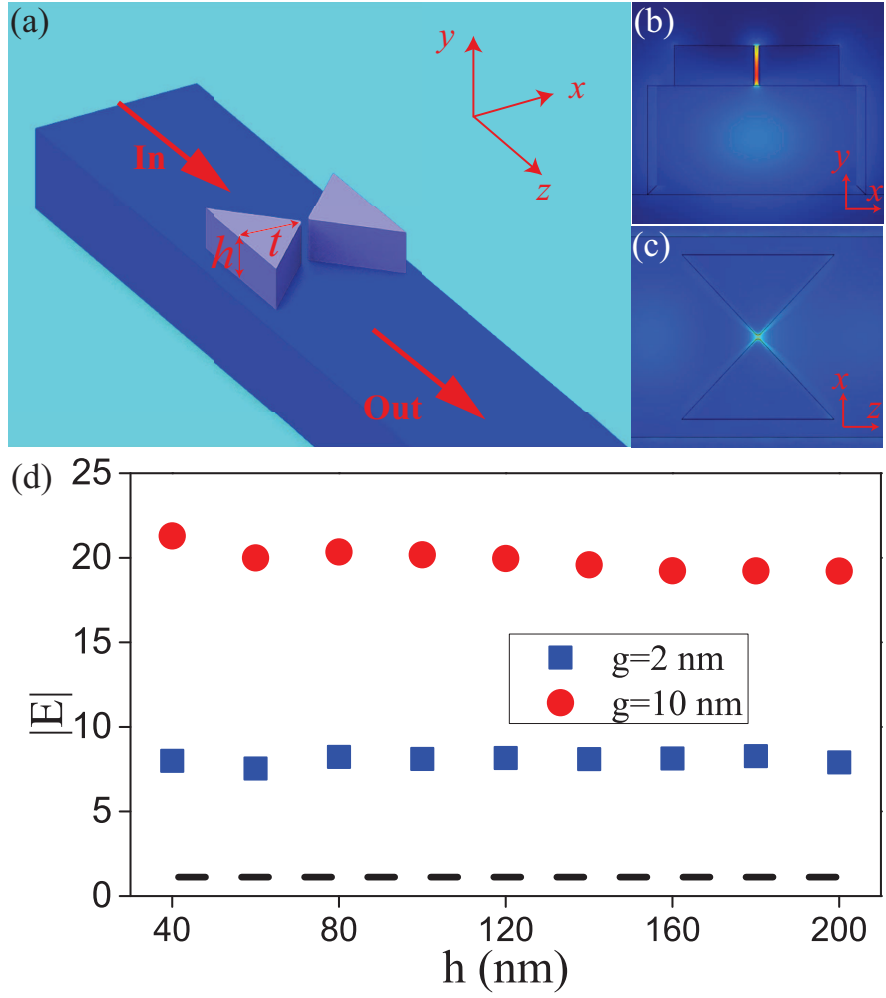}}
\caption{(a) Schematic of the dielectric bow-tie antenna. Side (b) and top
(c) views of the $|E|$ of the dielectric bow-tie antenna. (d) Maximum
of $|E|$ at the center of the gap for $g=2$ nm, $g=10$ nm, respectively,
as a function of $h$.}
\label{fig6}
\end{figure}

The corner and gap effects have been demonstrated above for dielectric
waveguides with different cross-section geometry, where light propagates
perpendicularly to the cross-section. Now, we turn to study the field
enhancement in the dielectric bow-tie (DBT) antenna structure, where
the corner and gap structures are not uniform along the waveguide.
This DBT antenna can be integrated with various photonic structures,
such as waveguide and microring, and is very potential for practical
applications.

As shown in Fig. 6(a), a dielectric bow-tie antenna is placed on a
silicon-on-insulator (SOI) waveguide with rectangle cross-section
of $450\,\mathrm{nm}\times250\,\mathrm{nm}$. The bow-tie antenna
consists of two opposing apex-to-apex silicon isosceles triangles,
the height and thickness of which are denoted as $h$ and $t$, respectively.
The vertex angle of the nano-triangle is $100{^\circ}$, and the radius
of the round is $r=10\,\mathrm{nm}$. Due to the effects of corner
and gap, we can expect a strongly enhanced field in the gap of the
DBT antenna. To investigate the field enhancement, the three-dimensional
model with $1550\,\mathrm{nm}$ fundamental TE-polarized mode loaded
in the waveguide is simulated numerically. From the field profiles
in Figs. 6(b) and 6(c), the electric field is greatly enhanced in
the gap of DBT antenna compared to that in the waveguide. Note that
the corner and gap effects can only be expected for the electric field
along the axis of symmetry of the corner, and there is no field enhancement
for the TM-polarized mode.

In Fig. 6(d), the maximum of $|E|$ at the center of the gap of DBT
antenna is plotted as a function of $h$ for $g=2\,\mathrm{nm}$ and
$g=10\,\mathrm{nm}$, respectively. The evanescent field of the fundamental
TE mode at the top surface of the waveguide without DBT antenna is
also shown by the dashed line. We can find that the maximum of $|E|$
is enhanced by $20$ and $8$ times for $g=2\,\mathrm{nm}$ and $g=10\,\mathrm{nm}$,
respectively. In both cases, the enhancement factor is insensitive
to $h$. The narrower gap leads to the stronger electric field enhancement,
which is consistent with the results in the apex-to-apex fan waveguides.
The DBT antenna is easy for fabrication and compatible with CMOS technology,
thus it will be useful as building blocks in integrated photonic circuits.

\section{Conclusion}

In summary, we demonstrate the corner- and gap-enhanced local electric
field in dielectric waveguide. As an example, in opposing apex-to-apex
fan-shaped waveguides with $r=g=10\,\mathrm{nm}$, the mode area of
fundamental wedge mode can be as small as $4\times10^{-3}A_{0}$ ($A_{0}=\lambda^{2}/4$),
far beyond the diffraction limit. The numerical results indicate that
the field enhancement and confinement abilities increase with decreasing
radius of rounded corner ($r$) and gap ($g$), and show singularity
for infinitesimal $r$ and $g$. The singularity behavior calls for
theoretical efforts to study the quantum effects of dielectric at
atomic scale. Furthermore, we propose a structure with dielectric
bow-tie antenna on a silicon-on-insulator waveguide, the field enhancement
of which is improved by one order. Although the waveguide studied
in this paper is focused at $1550$ nm, we should be aware that the
corner and gap effects are broadband. The lossless dielectric corner
and gap structures offer an alternative method to enhance the light-matter
interaction without metal nano-structure, and will find applications
in quantum electrodynamics, sensors and nano-particle trapping.

\section*{Acknowledgments}

We thank Xiao Xiong for discussion. CLZ is supported by the 973 Programs
(No. 2011CB921200). FJS is supported by the National Natural Science
Foundation of China (No. 11204169 and No. 11247289).



\end{document}